# RRAM-Based Analog Matrix Computing for Massive MIMO Signal Processing: A Review

**Authors:** Pushen Zuo[1,2] and Zhong Sun[1,2,3]*

**Affiliations:**

[1]Institute for Artificial Intelligence, Peking University, Beijing 100871, China

[2]School of Integrated Circuits, Peking University, Beijing 100871, China

[3]Beijing Advanced Innovation Center for Integrated Circuits, Beijing 100871, China

*Correspondence to: zhong.sun@pku.edu.cn

**Abstract**

Resistive random-access memory (RRAM) provides an excellent platform for analog matrix computing (AMC), enabling both matrix–vector multiplication (MVM) and the solution of matrix equations through open-loop and closed-loop circuit architectures. While RRAM-based AMC has been widely explored for accelerating neural networks, its application to signal processing in massive multiple-input multiple-output (MIMO) wireless communication is rapidly emerging as a promising direction. In this Review, we summarize recent advances in applying AMC to massive MIMO, including DFT/IDFT computation for OFDM modulation and demodulation using MVM circuits; MIMO detection and precoding using MVM-based iterative algorithms; and rapid one-step solutions enabled by matrix inversion (INV) and generalized inverse (GINV) circuits. We also highlight additional opportunities, such as AMC-based compressed-sensing recovery for channel estimation and eigenvalue circuits for leakage-based precoding. Finally, we outline key challenges—RRAM device reliability, analog circuit precision, array scalability, and data conversion bottlenecks—and discuss the opportunities for overcoming these barriers. With continued progress in device–circuit–algorithm co-design, RRAM-based AMC holds strong promise for delivering high-efficiency, high-reliability solutions to (ultra)massive MIMO signal processing in the 6G era.

## I INTRODUCTION

Massive Multiple-Input Multiple-Output (MIMO) has become a cornerstone technology in 5G-Advanced (5G-A) mobile networks [1] and is expected to further propel communication systems into the 6G era. Compared with conventional MIMO used in earlier generations, the number of antennas deployed at base stations (BSs) has increased to the order of hundreds [2], and the number of concurrently served users has grown to ten or more. At the same time, high-order quadrature amplitude modulation (QAM) schemes have been incorporated into modern communication standards—for instance, Wi-Fi 6 (802.11ax) supports 1024-QAM. These advancements collectively enhance wireless service quality, improving key performance indicators such as channel capacity, data rate, and energy efficiency [3].

In a massive MIMO system, a wide range of signal processing algorithms must be executed at the base station (BS) (Fig. 1a), and most of them inherently involve extensive matrix operations. For example, the discrete Fourier transform (DFT) and inverse DFT (IDFT)—used for OFDM modulation and demodulation—can be expressed as matrix-vector multiplication (MVM). Linear signal detectors and precoders typically require computing matrix inverses (INV) or generalized inverses (GINV). Other detection methods, such as belief propagation (BP) [4] and expectation propagation (EP) [5], also rely on matrix inversion within their algorithmic pipelines. Moreover, channel estimation can be formulated using compressed sensing techniques, which involve solving underdetermined linear systems with an $l_0$-norm constraint.

Matrix computations on digital processors are computationally expensive and typically exhibit polynomial-time complexity. For example, MVM for DFT and IDFT has a complexity of $O(N_s^2)$, where $N_s$ is the number of samples. Although fast algorithms—namely the fast Fourier transform (FFT) and inverse FFT (IFFT)—reduce this complexity to $O(N_s \log N_s)$, the computational burden remains substantial. Meanwhile, INV, which lies at the core of precoding, detection, and channel estimation, is even more demanding, with a complexity of $O(N^3)$ for an $N \times N$ matrix. For illustration, computing a GINV via Cholesky decomposition begins with two steps: matrix-matrix multiplication to form the Gram matrix, paralleled by an MVM operation. Subsequently, the Cholesky decomposition relies on the deployment of multiplier networks and adder trees to execute two-layer loops of massive multiply-accumulate (MAC) computations, exhibiting cubic computational complexity (Fig. 1b) [6]. Beyond these MAC operations, the decomposition also requires complex scalar computations—specifically division and square-root operations—along with basic vector additions and subtractions. The second stage of GINV is then computed through forward/backward substitution operations using the decomposed matrices, exhibiting quadratic complexity. As communication systems continue to scale, the computational load on baseband processors increases accordingly, posing significant challenges for designing high-speed signal processing hardware.

Although enhanced digital architectures have been proposed—such as designs that reduce the number of clock cycles compared with conventional implementations or employ multi-core parallelism for FFT acceleration [7]—their achievable throughput remains limited to only a few Gb/s. This is insufficient for the data-rate requirements of 5G-A and emerging 6G base stations, which are expected to range from 10 Gb/s to 100 Gb/s [8]. The root cause lies in the fundamental characteristics of digital computing: arithmetic operations are executed sequentially, and each operation must be constructed using resource-intensive Boolean logic gates operating on binary data. These inherent constraints prevent any fundamental reduction in computational complexity, thereby limiting further gains in transmission speed. From an energy standpoint, solving a linear precoding or detection problem for an 8×128 MIMO system at 6G data rates is estimated to consume approximately 10 W—assuming an energy efficiency of 100 pJ/bit—which is prohibitively high for practical BS deployment. As matrix dimensions continue to grow, the associated energy consumption is expected to increase even more rapidly.

As the challenges associated with advanced technology nodes continue to intensify, relying solely on process scaling to enhance the performance of MIMO baseband processors is becoming unsustainable for meeting the stringent requirements of 6G communication systems. Moreover, the von Neumann architecture underlying digital hardware imposes substantial limitations in data-intensive applications, where frequent data movement between the processor and memory incurs significant time and energy overhead. As a result, a new computing paradigm is needed to bridge the growing gap between the slowing pace of hardware improvement and the rapidly increasing performance demands of advanced signal processing algorithms.

Recently, RRAM-based analog matrix computing (AMC) has emerged as a fast and energy-efficient approach for performing matrix operations [9]. In AMC systems [10], matrix values are encoded directly as conductance states within an RRAM array (Fig. 1c), which typically adopts a 1-Transistor-1-resistor (1T1R) cell structure to mitigate sneak-path currents. The RRAM devices themselves inherently support AMC with several advantages: a simple structure consisting only of a top electrode (TE), bottom electrode (BE), and resistive switching layer; analog-tunable conductance; and low-power read/write characteristics. They also offer high cross-point integration density and strong compatibility with CMOS technology, facilitating seamless integration with peripheral circuits. The AMC architecture enables massive spatial parallelism, leading to substantial improvements in both throughput and energy efficiency. Furthermore, AMC performs computation directly within memory (Fig. 1d), thereby eliminating the memory–processor bottleneck that constrains digital hardware and further reducing latency and power consumption [11]. Fundamental matrix operations—including MVM [12], INV [13], GINV [14], and eigenvector computation [13]—have already been demonstrated in hardware, often with theoretical time complexity approaching $O(1)$ [15–17]. These capabilities have enabled AMC to accelerate a variety of applications in machine learning [14, 18–20], scientific computing [21–23], and graph algorithms such as PageRank [13].

Massive MIMO signal processing is emerging as a promising application area for RRAM-based AMC. As 6G technologies push antenna densities even higher [24], the number of antennas at a base station is expected to scale to the order of 1000 [25], with the number of simultaneous users reaching around 100. These trends produce problem sizes that are large, yet still well within the manageable range of AMC circuits. Moreover, wireless channels are typically modeled as noisy systems, meaning that many associated signal processing tasks are inherently error-tolerant, thereby relaxing the precision requirements on AMC. In addition, the matrices involved in inversion are often Gram matrices derived from the wireless channel, which are generally well-conditioned—further supporting the suitability of AMC-based solutions. Collectively, these factors position RRAM-based AMC as a compelling candidate for accelerating massive MIMO signal processing.

In this Review, we present a comprehensive overview of AMC-based massive MIMO signal processors. We begin by introducing fundamental matrix-operation circuits enabled by AMC, highlighting their inherent one-step computing capability. Next, we discuss the application of AMC to various signal processing algorithms in massive MIMO systems. Finally, we outline key challenges and potential solutions for advancing AMC-based MIMO signal processing hardware.

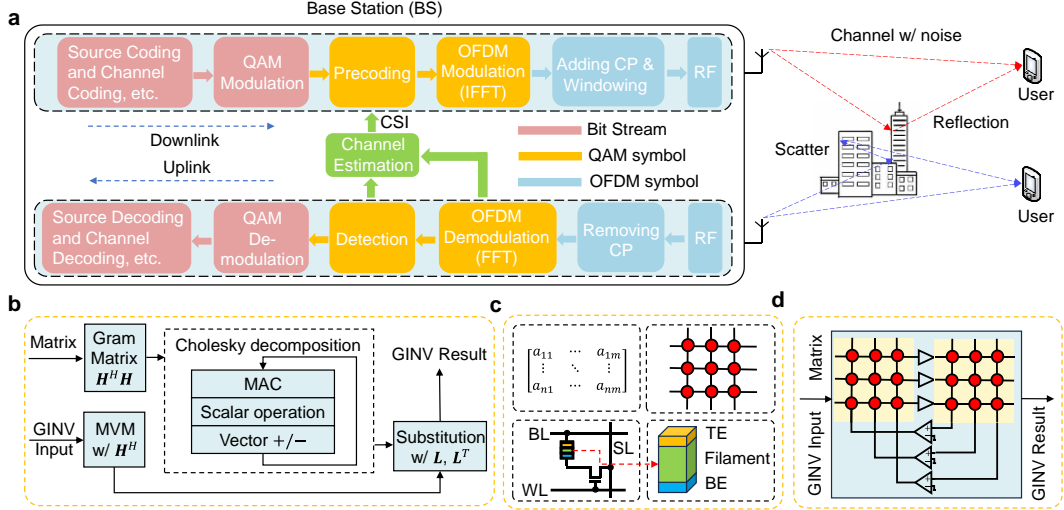

**Fig. 1. RRAM-Based AMC for signal processing in BS.** (a) Data flow and signal processing algorithms in the BS. In the downlink, after source and channel coding followed by QAM modulation, the baseband processing unit performs matrix-based operations such as precoding and OFDM modulation, using channel state information (CSI) obtained from channel estimation. After cyclic prefix (CP) insertion and windowing, the processed signal is transmitted to users through the RF front-end. The uplink follows the reverse sequence, including CP removal, OFDM demodulation, detection, and decoding. In practical wireless environments, scattering and reflection determine the channel matrix, and channel estimation itself requires matrix computations. (b) Hardware architecture of a digital GINV unit (the core operation of MMSE detection) based on Cholesky decomposition. Here, $H$ is the channel matrix and $L$ is the decomposition result. (c) A matrix can be directly mapped onto an RRAM array composed of 1T1R cells integrating resistive-switching devices. The matrix elements are encoded by programming the conductance states of these RRAM devices. (d) Hardware architecture of AMC-based MMSE detection. After the input matrix is programmed into the RRAM array, the GINV circuit directly produces the detection result once the input vector is applied—eliminating the multi-stage pipelined matrix/vector/scalar operations required in (b).

## II AMC CIRCUITS

Fig. 2a illustrates an AMC circuit for performing MVM, i.e., $y = Ab$, where $A$ is the matrix, $b$ is the input vector, and $y$ is the output vector. Each matrix element is represented by the analog conductance of the corresponding RRAM cell in the array, while the input vector is applied as a set of analog voltages. According to Ohm's law and Kirchhoff's current law (KCL), the output currents collected on the column lines form the resulting output vector. The most direct readout method uses a transimpedance amplifier (TIA) to convert the column currents into voltages. To interface with digital systems, an analog-to-digital converter (ADC) is typically added. To support arbitrary real-valued matrices with both positive and negative elements, common techniques include column-splitting and row-splitting using operational-amplifier (OPA)-based analog inverters. To avoid the high power and area overhead of analog inverters, a conductance compensation strategy has been proposed, in which compensating resistive elements are added to ensure equal total conductance in each pair of rows. The two rows are then connected to the inverting and non-inverting inputs of a TIA, which inherently performs subtraction. Although TIA-based designs are intuitive and widely used, their high power consumption and large footprint motivate the search for more efficient alternatives. Depending on the computational domain, MVM peripheral circuits can operate in the current [26], voltage [27], charge [28], or time domains [29], enabling the use of compact and energy-efficient analog circuits such as sense amplifiers and time-to-digital converters.

INV can be viewed as the reverse operation of MVM, and its implementation using an RRAM array is illustrated in Fig. 2b. The circuit components for INV are largely identical to those used in MVM; the key difference lies in the circuit topology, which determines the circuit's function. Specifically, the matrix $A$ is stored in the RRAM array, as in MVM. A set of analog voltage signals is applied across fixed resistors $G_0$, connected to the row lines of the array, representing the input vector $b$. The output vector $x$, generated by a set of OPAs, is fed back to the column lines, forming a negative feedback loop. According to circuit physics and OPA feedback principle, the system stabilizes at a state that satisfies $Ax = b$. At equilibrium, the output voltage vector corresponds to the solution $x = A^{-1}b$. To support real-valued matrices in INV, techniques such as column-splitting, row-splitting, and conductance compensation-assisted row-splitting—previously used for MVM—can also be applied [30]. For circuit stability, the matrix $A$ must be positive definite [15], a condition typically satisfied

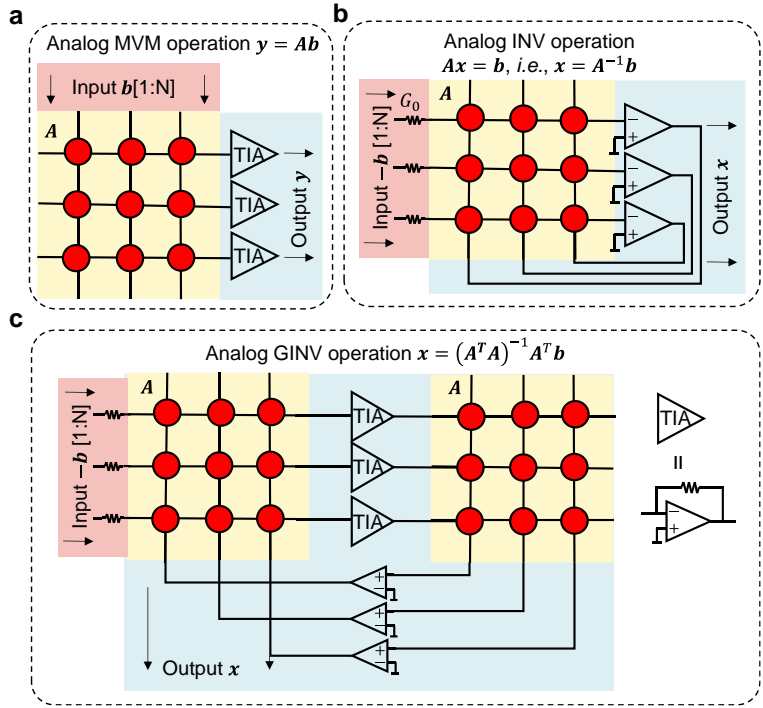

**Fig. 2. AMC circuits for various matrix operations.** (a) MVM. (b) INV. (c) GINV. The circuit components used across these matrix operations are nearly identical; the specific circuit-connection topology determines which operation is performed.

in massive MIMO signal processing, as discussed later.

To solve $Ax = b$ when $A$ is a non-square matrix, different approaches are needed depending on whether the system is overdetermined or underdetermined. For instance, MIMO detection in wireless communication is typically an overdetermined problem, where $A \in R^{n \times m}$ with $n>m$. This problem is usually solved by minimizing the L$_2$-norm $\|b - Ax\|_2$, resulting in a solution based on the left pseudoinverse: $x = (A^T A)^{-1} A^T b$. Conversely, MIMO precoding often leads to underdetermined problems, where $A \in R^{n \times m}$ with $n<m$. These problems are typically solved under the constraint of minimizing $\|x\|_2$, giving the right pseudoinverse solution: $x = A^T (AA^T)^{-1} b$. Fig. 2c illustrates a GINV circuit for computing the left pseudoinverse, consisting of two identical RRAM arrays, a set of TIAs, and a second set of positive-feedback OPAs. The known vector $b$ is applied to the rows of the left RRAM array. The same circuit structure can be used for computing the right pseudoinverse by storing the transposed matrix $A^T$ in the arrays and applying the input vector to the columns of the right array. It is worth noting that the GINV circuit can also compute the inverse of a square matrix, especially when the matrix is not positive definite—cases that cannot be handled by the single-array INV circuit shown in Fig. 2b. This circuit has been proven to be Lyapunov stable for any given matrix $A$ [17].

## III AMC Circuits For Massive MIMO Signal Processing

Given that massive MIMO signal processing inherently involves a large volume of diverse matrix computations, the application of AMC technology for computational acceleration and energy savings at the base station is both natural and promising. In particular, as advanced MIMO systems scale to ultra-large dimensions—ranging from 256×256 to 1024×1024—RRAM-based AMC demonstrates strong potential to deliver the extremely high throughput required, which would otherwise pose significant challenges for conventional digital processors. In the following sections, we describe AMC-based algorithm implementations, including DFT/IDFT, detection, precoding, and channel estimation, and summarize recent progress in this area.

### A. MVM circuit for massive MIMO signal processing

MVM serves as a fundamental operation in matrix computing, making AMC-based MVM circuits highly suitable for widespread use in computational acceleration. Depending on algorithmic requirements, AMC-based MVM can be combined with other operations to enhance the performance of related applications. A notable example over the past decade is their use in neural network acceleration, which has attracted significant attention [31-33]. In massive MIMO signal processing, several essential operations are inherently MVM in nature, and AMC circuits can similarly be leveraged to accelerate these operations within advanced algorithms. Typical applications of AMC-based MVM circuits include DFT/IDFT, OFDM modulation/demodulation, and hybrid precoding and detection.

**DFT/IDFT.** The DFT and IDFT are fundamentally MVM operations performed in the complex domain [34–36]. To implement them using AMC-based real-valued MVM circuits (Fig. 3a), the complex DFT matrix $W_{DFT}$ and input/output vectors $x_{DFT}$ and $y_{DFT}$ are decomposed into real and imaginary parts. For example, the real-valued representation of the complex DFT matrix is: $\begin{bmatrix} W_{DFT,R} & -W_{DFT,I} \\ W_{DFT,I} & W_{DFT,R} \end{bmatrix}$, where $W_{DFT,R}$ and $W_{DFT,I}$ denote the real and imaginary components, respectively. Similarly, the real-valued representations of $x_{DFT}$ and $y_{DFT}$ are formed by stacking their real and imaginary parts. Each matrix element can be directly mapped to an analog device or distributed across several single-bit or multi-bit devices using bit-slicing [21, 37-39]. The input vector is converted into a set of input voltages for the AMC circuit, and the resulting output analog signals are digitized by ADCs. Since the IDFT matrix is the conjugate transpose of the DFT matrix, its implementation follows the same principle. AMC-based DFT/IDFT execution has demonstrated up to two orders of magnitude improvement in computational efficiency over conventional digital implementations in BS signal processing hardware [34].

**OFDM demodulation/modulation.** While stand-alone AMC-based DFT/IDFT units can accelerate frequency-domain transformation, they still necessitate ADC/DAC interfaces and additional steps such as cyclic prefix (CP) removal to connect with RF modules [34–36]. To mitigate this overhead, recent systems integrate OFDM processing directly with analog computation and RF front-end modules, enabling end-to-end analog-domain OFDM transceiver functionality [40].

As shown in Fig. 3b, on the transmitter side, the hardware generates orthogonal subcarrier signals—$\cos(k\omega t)$ and $-\sin(k\omega t)$ (where $\omega$ is the base frequency, $t$ is time, and $k = 1,2,...$)—and performs OFDM modulation entirely in the analog domain using these subcarriers. Both the generation of these subcarriers and the modulation operation are realized on RRAM arrays by programming modulation coefficients into the array. Conceptually, the orthogonal subcarriers are produced by multiplying the IDFT matrix with a sequence of one-hot encoded vectors applied periodically as voltage signals. Subsequently, the orthogonal subcarriers are fed in parallel into the array—programmed with transmitted digital data—to facilitate OFDM modulation. At the receiver, following RF down-conversion and low-pass filtering (LPF), the sampled time-domain waveform is multiplied by the DFT matrix to recover frequency-domain subcarrier symbols, which are then mapped to constellation points using analog hard-decision logic. This fully analog implementation—covering down-conversion, mixing, LPF, and DFT-based demodulation — eliminates intermediate AD/DA conversions and allows seamless integration with RF modules. In particular, mixing functions to multiply the received high-frequency RF signal with a local oscillator (LO) signal, translating the RF signal to an intermediate frequency (IF) or baseband (BB) for easier subsequent processing. This mixing operation can likewise be implemented using AMC circuits. For instance, analog down-conversion can be realized by storing the LO signal (*e.g.,* carrier-synchronized signal) in an AMC matrix and applying the RF signal from the front-end as the input [36]. The LPF stage can be represented as a Toeplitz matrix, enabling time-domain convolution through analog MVM.

**Hybrid precoding.** Beyond immediate applications such as DFT and OFDM demodulation/modulation, MVM circuits can also accelerate a wide range of matrix-computation-based algorithms, similar to their use in neural network acceleration. Hybrid precoding is a technique that converts digital baseband signals into analog RF signals and is used for spatial-domain pre-processing in millimeter-wave communication system. It adjusts the amplitude and phase of each antenna in the array by multiplying the digital signal with the precoding matrix, which consists of both digital and analog precoders. The core task in hybrid precoding is obtaining the analog precoder, which involves solving a non-convex optimization problem. This problem can be addressed using machine learning methods, which create hardware bottlenecks on conventional digital computers as the matrix size increases. AMC-based approaches were initially proposed using a self-attention-based unsupervised neural network [41] (Fig. 3c). The implementation leverages RRAM arrays to perform linear transformations and convolution operations through in-memory MVMs. To mitigate IR-drop effects caused by wire resistance and access resistance in large-scale RRAM arrays, a quantization-aware compensation model was developed and integrated with dynamic scaling of MVM input vectors, effectively reducing computational errors induced by non-ideal circuit characteristics.

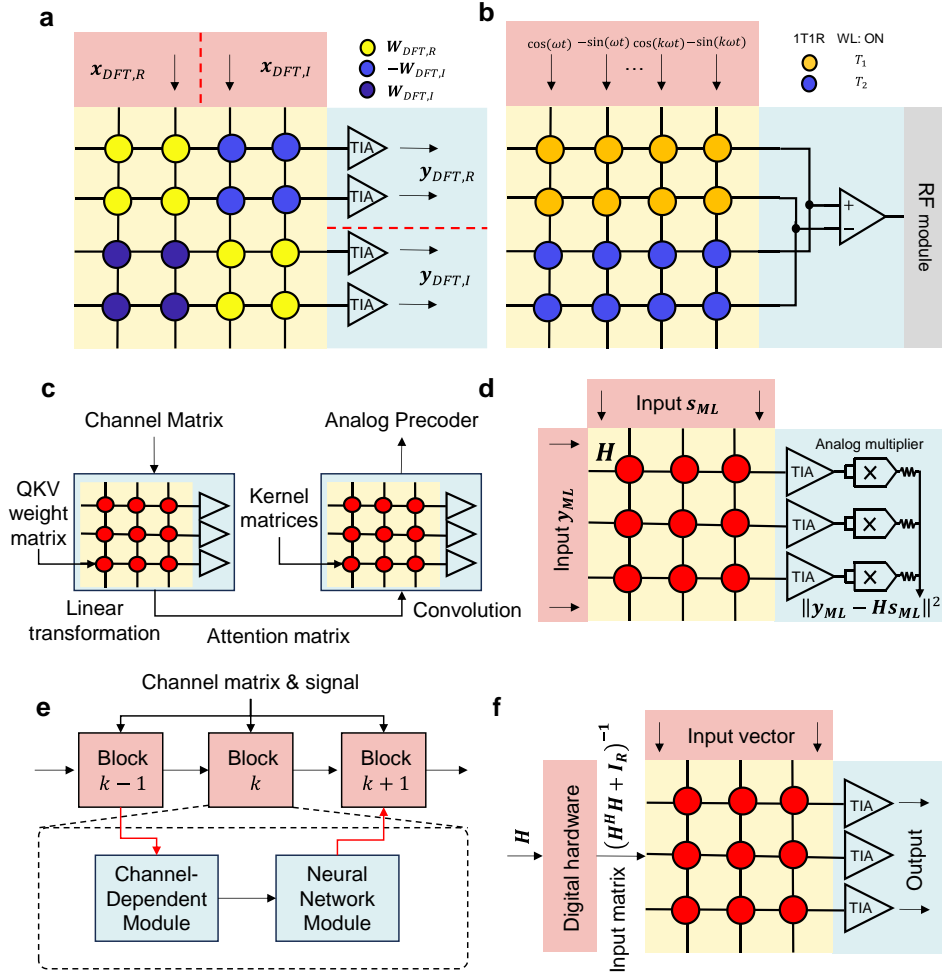

**Fig. 3. AMC MVM circuit applications in massive MIMO algorithms.** (a) Complex-domain AMC MVM circuit. Complex-valued MVM operations are realized by decomposing complex matrices and vectors into their real and imaginary components, which are then mapped onto the AMC circuit. (b) DAC-free wireless transmitter architecture. OFDM modulation is achieved using multiple orthogonal signals and RRAM arrays preprogrammed with transmission data, eliminating the need for DACs in the RF front-end. (c) Hybrid precoding architecture using AMC circuits. The structure contains two core computational components—both implemented using MVMs: linear transformations between the input channel matrix and the trainable Query, Key, and Value (QKV) weight matrices, and convolution operations between the input attention matrix and the trainable kernel matrices. The trainable matrices are programmed into the array. (d) AMC circuit for accelerating ML detection, which requires exhaustive evaluation of all candidate transmitted symbol vectors $s_{ML}$, performing MVMs and analog multiplications in computing $\|y_{ML} - Hs_{ML}\|^2$. (e) Deep-unfolding–based MIMO detector, where both the channel-dependent module and the channel-independent neural network in each block are implemented using AMC MVM. (f) AMC MVM-based MIMO detection, where the programmed inversed matrix $(H^H H + I_R)^{-1}$ is precomputed using digital hardware and then programmed into the array.

**Detection.** In many MIMO signal processing scenarios, such as precoding and detection, computation can be accelerated by performing analog MVM operations within discrete algorithms. For example, to accelerate maximum-likelihood (ML) detection—a theoretically optimal algorithm defined as $\boldsymbol{x}_{ML} = \arg\min_{s}\|\boldsymbol{y}_{ML} - \boldsymbol{H}\boldsymbol{s}_{ML}\|^2$—AMC approaches can be employed [42] (Fig. 3d). In this setup, MVM circuits compute $\boldsymbol{H}\boldsymbol{s}_{ML}$, while analog scalar multipliers and MAC units handle the squared norm $\|\boldsymbol{y}_{ML} - \boldsymbol{H}\boldsymbol{s}_{ML}\|^2$. Here, $\boldsymbol{y}_{ML}$ denotes the received current signals, $\boldsymbol{H}$ is the channel matrix, $\boldsymbol{s}_{ML}$ is the estimated voltage signal, and $\boldsymbol{x}_{ML}$ is the detection result. While the MVM component is accelerated, the complexity of enumerating $\boldsymbol{s}_{ML}$ remains exponential, as all possible candidates must be evaluated. This limitation has motivated the development of deep MIMO detectors, which closely approach optimal performance. These can be constructed by integrating an analog MVM-based channel-dependent module with a neural network module within each block [43] (Fig. 3e). This architecture not only adapts to channel variations but also compensates for non-ideal effects such as programming errors through neural network training. Although online inference for MIMO detection achieves $O(1)$ complexity, the offline training process remains both data- and computation-intensive.

Matrix multiplication serves as the cornerstone of an alternative approach that employs near-optimal linear methods, whose mathematical essence resides in higher-complexity linear algebraic computations—specifically, INV and GINV—which are traditionally executed on digital computers using techniques such as the Neumann series [44–45], QR decomposition [6, 46], and Gauss-Jordan elimination [47]. These techniques break the computation into multiple steps of MVMs or a series of vector and scalar operations. In the context of AMC-based MIMO signal processing hardware, initial studies of linear detection pre-computed the inverse matrix $(\boldsymbol{H}^H\boldsymbol{H} + \boldsymbol{I}_R)^{-1}$ using digital hardware, where $\boldsymbol{I}_R$ is the regularization term (*e.g.*, $\boldsymbol{I}_R = \boldsymbol{0}$ for the zero-forcing algorithm and $\boldsymbol{I}_R = 1/\text{SNR}$ for the minimum mean square error (MMSE) algorithm, with SNR denoting the signal-to-noise ratio). The resulting inverse matrix was then used with RRAM arrays to perform MVMs [48] (Fig. 3f). However, this approach does not address the inherent computational bottleneck of performing INV operations on conventional digital hardware. In some studies, the computation of GINV has been reformulated as a gradient descent problem to solve for the optimal pseudoinverse matrix [49]. Although this method reduces the computational complexity to $O(N)$ compared with pre-computation, the continuous reprogramming of devices to implement matrix updates introduces significant overhead in both computational latency and energy consumption.

### B. INV/GINV circuits for precoding and detection

Massive MIMO signal processing involves extensive matrix inversion operations. Beyond using AMC circuits to accelerate MVM within iterative solving algorithms, a more direct approach is to employ AMC-based INV or GINV circuits for signal processing tasks—primarily precoding and detection—where linear methods correspond to exact INV or GINV operations, potentially with certain regularization terms. A pioneering study implemented linear zero-forcing (ZF) precoding using AMC circuits [50], where the GINV computation is decomposed into two parts: INV and MVM. Both components are realized with AMC circuits and connected via analog voltage followers to enable continuous analog data flow (Fig. 4a). Leveraging the one-step operation capability of AMC-based INV and MVM circuits, along with optimized OPAs, the design achieves ZF precoding for a 16×128 MIMO system in 20 ns in simulation. This results in approximately a 50× improvement in energy efficiency but a 2 dB SNR loss at a bit error rate (BER) of $10^{-3}$ compared with conventional digital hardware. The matrix GINV can also be computed in one step using an AMC-based GINV circuit [34], enabling the implementation of two linear detection algorithms: MMSE and ZF. In Fig. 4b, the complex-valued channel matrix $H$ is decomposed as $\begin{bmatrix} Re(H^+) - Re(H^-) & Im(H^-) - Im(H^+) \\ Im(H^+) - Im(H^-) & Re(H^+) - Re(H^-) \end{bmatrix}$, where $Re(H^+)$, $Re(H^-)$, $Im(H^+)$, and $Im(H^-)$ denote the positive/negative and real/imaginary parts, respectively. Due to this decomposition, the setup is limited to 4×4 MIMO linear detection, exhibiting an SNR loss of approximately 5 dB at a BER of $10^{-3}$ compared with conventional digital hardware.

In addition to ZF and MMSE, AMC methods enable the implementation of algorithms designed to address more complex communication challenges. Compared with ZF, regularized ZF (RZF), expressed as $x_{RZF} = (H^H H + \lambda I)^{-1} H^H u_{RZF}$, incorporates an additional regularization term $\lambda I$, analogous to that in MMSE. Here, $x_{RZF}$ and $u_{RZF}$ denote the output and input vectors of RZF, respectively. RZF computation can be realized by modifying the GINV circuit to implement a ridge regression circuit [51–53]. Notably, to minimize circuit power consumption and complexity, an analog inverter-free design shifts the conductance values upward, converting both positive and negative matrix elements into positive conductance values [53]. The ridge regression circuit is further optimized: a design that originally required two arrays can be implemented using a single array (Fig. 4c). Moreover, considering the disparity between large-scale fading coefficients (LSFC) and small-scale fading coefficients (SSFC) in MIMO detection [54–55], ZF and MMSE algorithms are modified as $x_{LSFC} = \Lambda^{-1}(G^T G + P)^{-1} G^T u_{LSFC}$, according to the channel matrix decomposition $H = G\Lambda$, where $G$ is the SSFC matrix, $\Lambda$ is the diagonal LSFC matrix, and $P$ is a diagonal matrix ($P = 0$ in ZF detection) associated with the average symbol energy of the transmitted signals and LSFCs between the users and the BS. These algorithms can be mapped to AMC-based detector circuits, including optimized GINV or decomposition circuits for $(G^T G + P)^{-1} G^T u_{LSFC}$ and amplifier-enhanced circuit for $\Lambda^{-1}$. A similar circuit architecture can also be extended to implement MIMO detection based on the alternating direction method of multipliers (ADMM) by incorporating analog adder-subtractor and comparator circuits [56].

Additionally, to improve detection performance beyond linear methods, some studies have proposed successive interference cancellation (SIC) detection, such as MMSE-SIC, implemented using AMC-based approaches [57]. MMSE-SIC detection requires cyclic execution of four steps: MMSE detection, slicing, interference cancellation, and matrix dimension reduction. In step 1, for MMSE detection, the AMC circuit computes $\boldsymbol{b}_k = \left(\boldsymbol{G}_{<k>}^H \boldsymbol{G}_{<k>} + \sigma_n^2 \boldsymbol{I}\right)^{-1} \boldsymbol{G}_{<k>}^H \left(\boldsymbol{y}_{SIC} - \boldsymbol{G}_{(k-1)} \boldsymbol{e}_{(k-1)}\right)$, which represents a fusion and improvement of the GINV and MVM circuits (Fig. 4d). Here, $\boldsymbol{b}_k$ is the resulting vector, $\boldsymbol{G}_{<k>}$ is the reordered transfer matrix from the $k$th to the $K$th symbol, $\boldsymbol{G}_{(k)}$ is the matrix for the same range, $\sigma_n^2 \boldsymbol{I}$ is the expected covariance matrix of zero-mean additive white Gaussian noise (AWGN), $\boldsymbol{y}_{SIC}$ is the received signal, and $\boldsymbol{e}_{(k)}$ denotes the estimated symbols from the first to the $k$th symbol. In step 2, for slicing, a hybrid analog-digital structure is proposed, where the analog output from the AMC detection circuit is converted into discrete symbols using voltage comparators, combinational logic circuits, and related components. During steps 3 and 4, the detected symbols are subtracted from the received signals (interference cancellation), and the matrix dimensions are reduced accordingly. As a result, the computational load of the AMC circuits decreases progressively with each detection iteration.

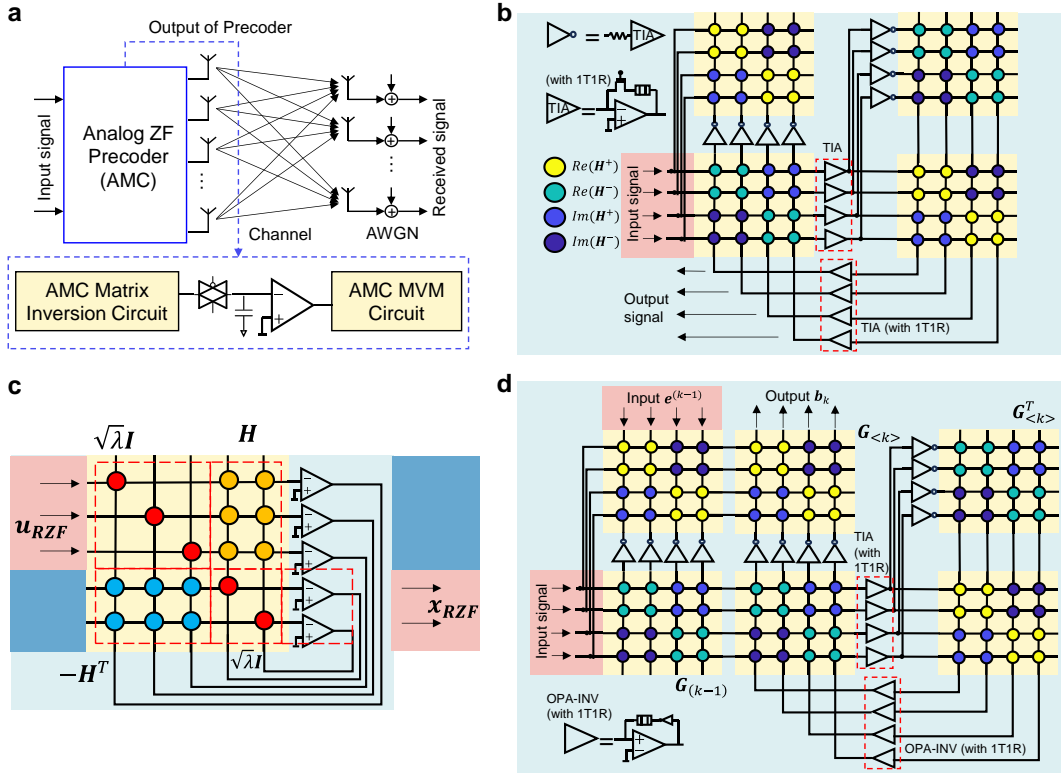

**Fig. 4. AMC applications in linear precoding and detection algorithms.** (a) AMC based ZF precoding circuit, consisting of an AMC INV unit followed by an MVM unit to compute the ZF precoder $\boldsymbol{H}^H(\boldsymbol{H}\boldsymbol{H}^H)^{-1}$ in two steps. A set of analog voltage followers connects the two circuits, enabling fully analog data flow. (b) AMC based L-MMSE and ZF detection circuit, with the GINV unit as the core. The processing of complex-valued matrices and vectors follows the same method shown in Fig. 3a. To adapt to different SNR conditions, the OPA feedback conductance is tuned using a bank of 1T1R devices. (c) AMC RZF circuit, implemented by converting the ridge-regression formulation into an inverting-amplifier–only architecture through block-matrix mapping. (d) AMC-based MMSE detection integrated into the MMSE-SIC framework. MMSE detection forms the first stage, after which the output is sliced, interference-canceled, and dimension-reduced to produce updated input signals. These refined signals are then fed back to the AMC circuit to detect the remaining symbols.

Considering interference among multi-user transmitters (Tx), multi-user receivers (Rx), and multi-antenna relays in MIMO communication, interference alignment (IA) is required. The opposite directional interference alignment (ODIA) algorithm, known for its strong performance, involves designing the relay beamformer $T$, which is conceptually similar to

**Table 1. Demonstrations of MIMO signal processing with AMC.**

| Reference | 34 | 36 | 40 | 41 | 42 | 43 | 49 |
|---|---|---|---|---|---|---|---|
| Signal processing | DFT/Detection | DFT | DFT | Hybrid precoding | Detection | Detection | Detection |
| Algorithm | MMSE (detection) | —— | —— | Self-attention | ML | Deep-unfolding | MMSE MMSE-SIC |
| Method | GINV | RF | RF | MVM | MVM | MVM | MVM |
| Configuration | 4×4 antennas | 64-point[③] | 2×2 antennas 15-point | —— | 4×64 antennas | 20×30 antennas | 32×4 antennas |
| Modulation | 4-QAM | 16-PSK | 4-QAM | —— | 16-QAM | 16-QAM | QPSK |
| Experiment | Yes | Yes | Yes | Yes | No | No | No |
| Throughput | 160.8 Gb/s[①] | 0.254 TOPS[④] | —— | —— | —— | 2.98 TFLOPS | —— |
| Energy Efficiency | 0.2 pJ/b[②] | 21.3 TOPS/W[⑤] | 222 TOPS/W | 1.545 TOPS/W | 1.248 TFLOPS/J[⑥] | —— | —— |

| Reference | 50 | 53 | 54 | 56 | 57 | 58 | 74 |
|---|---|---|---|---|---|---|---|
| Signal processing | Precoding | Precoding/Detection | Detection | Detection | Detection | ODIA | Detection |
| Algorithm | ZF | RZF | ZF/MMSE | ADMM | MMSE-SIC | —— | ZF |
| Method | INV+MVM | INV | INV | GINV | GINV | INV+MVM | INV |
| Configuration | 128×16 antennas | 10×5 antennas | 4×64 antennas | 64×72 antennas | 32×64 antennas | Up to 32×32 mat. | 128×8 antennas |
| Modulation | 16-QAM | 32-QAM | 64-QAM | 64-QAM | 16-QAM | —— | 256-QAM |
| Experiment | No | Yes | No | No | No | No | Yes |
| Throughput | 20 ns/mat.[⑦] | ~0.1 TOPS | —— | 13.8 TFLOPS | 5.5 TOPS | —— | Up to 2 TOPS |
| Energy Efficiency | 2.5 nJ/mat. | ~1 TOPS/W | 2~20 TOPS/W | 5.18 TFLOPS/J | 1.41 TOPS/W | —— | Up to 6 TOPS/W |

① Gb/s: one billion (Giga) bits of data transferred per second. ② pJ/b: picojoules ($10^{-12}$ J) of energy consumed per transmitted or received bit. ③ 64-point: 64 discrete time-domain or frequency-domain samples used for domain conversion. ④ TOPS: tera operations per second. ⑤ TOPS/W: tera operations per second per watt. ⑥ TFLOPS/J: tera floating-point operations per second per joule. ⑦ ns/mat.: nanoseconds per matrix operation, *i.e.,* the time required to complete one matrix operation.

precoding [58]. The problem can be formulated as $\boldsymbol{H}^{RB}\boldsymbol{T}\boldsymbol{H}^{UR} = -\boldsymbol{H}^{UB}$, where $\boldsymbol{H}^{RB}$, $\boldsymbol{H}^{UR}$, and $\boldsymbol{H}^{UB}$ denote the channel matrices between the relay and Rx, Tx and relay, and Tx and Rx, respectively, and $\boldsymbol{T}$ is the relay beamforming matrix to be solved. ODIA is implemented through a series of INV and matrix multiplication operations, which can be accelerated using AMC. Simulation results report improvements in power efficiency by tens of times. The parameters and performance of the aforementioned works are summarized in Table 1.

## IV CHALLENGES AND OPPORTUNITIES

AMC holds significant potential for accelerating massive MIMO signal processing in hardware, improving performance metrics such as data rate and energy efficiency. However, several key challenges remain in applying AMC circuits to BS systems. At the device level, reliability issues must be addressed, while at the circuit level, considerations such as power efficiency require careful attention. Furthermore, algorithm design must compensate for the inherent limitations of analog circuits, including low precision and small scale, to support high-precision, large-scale, and diverse computational tasks. Additionally, integration between AMC circuits and other BS modules (e.g., RF chains) remains insufficiently explored. More algorithms tailored to various signal processing stages and channel conditions need to be implemented on AMC hardware. An overview of the challenges and opportunities for AMC is presented in Fig. 5a.

### A. Limitations of RRAM devices and analog circuit

The deployment of RRAM-based AMC in massive MIMO systems faces several critical device-level challenges. One primary obstacle is that rapidly changing channel conditions require real-time matrix reconfiguration due to the dynamic nature of wireless communication environments. This imposes stringent demands on the write/erase speed of RRAM devices. For programming RRAM devices to multiple levels in general AMC circuits, the write-verify method is a reliable approach [59–60]. However, this methodology requires multiple cycles to achieve the target states, which inevitably compromises programming speed and increases energy consumption. This highlights the need for more efficient solutions, such as fully analog one-step programming schemes.

Furthermore, the limited endurance of RRAM conflicts with the continuous parameter updates required for precoding and signal detection in time-varying channels. From a materials perspective, endurance can be improved through process techniques, such as employing high-barrier metals (e.g., Ti and Ta) as electrodes [61, 62] or introducing buffer layers to increase the potential barrier for oxygen vacancy migration [63]. From a programming perspective, optimized waveforms—such as triangular ramped pulses—can mitigate current overshoot during RESET operations, reducing device stress and improving endurance [64]. At the system level, error correction codes (ECC) provide an effective solution for programming failures in RRAM arrays and have been widely applied in RRAM-based in-memory computing [65, 66]. Additionally, fault-tolerant DFT and MIMO detection can be achieved for defective devices through a combination of software-based matrix decomposition and hardware-level compensatory design [67].

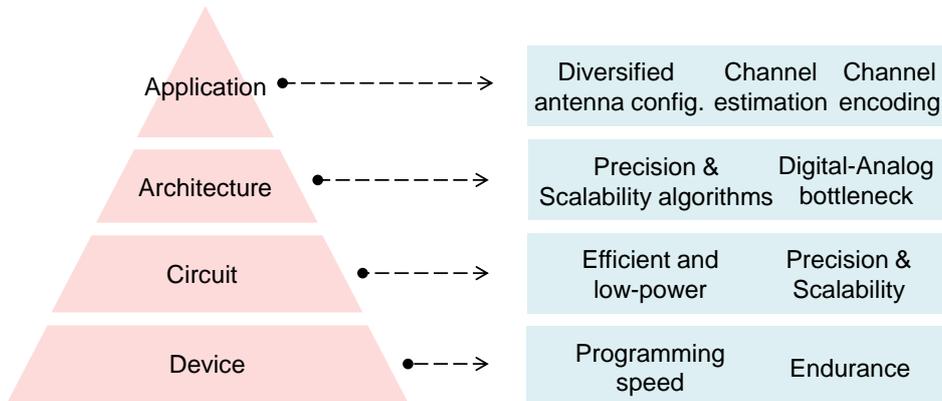

**Fig. 5. The challenges for AMC Massive MIMO hardware implementations.** The challenges are grouped into four system levels according to the order of the system from bottom to top.

For AMC circuits in massive MIMO, efficient and low-power computation is a key optimization goal. In these circuits, most of the power consumed in matrix computations originates from OPAs and ADCs [55, 68–70]. Consequently, the power requirement of analog circuitry scales at least linearly with matrix size, underscoring the need for more energy-efficient designs. Within the practical programming precision range of RRAM arrays (e.g., 2–4 bits), there is substantial redundancy in the open-loop gain of OPAs [54]; for example, achieving an error of ~10% requires only a 40 dB gain. Similarly, excessive precision in ADCs and digital-to-analog converters (DACs) introduces unnecessary power overhead. Therefore, future research should focus on optimizing AMC parameters based on realistic computational requirements—for instance, by developing simulation frameworks for MIMO precoding [71]—and adopting analog circuit architectures better aligned with MIMO applications.

### B. Precision issue of AMC

The SNR loss discussed in Section III reflects a longstanding limitation in AMC research: low computational precision. Precision constraints arise from several factors. First, analog circuits inherently exhibit higher noise than digital circuits. Second, the storage precision of RRAM devices further limits computational accuracy, as the actual programmed conductance is affected by programming variability, retention time, and I–V nonlinearity. Third, as matrix sizes increase, line resistance and parasitic effects in RRAM arrays degrade output precision.

In MVM operations, these limitations can often be mitigated using bit-slicing and analog-slicing techniques [21–22, 37–39], since the distributive property allows decomposition of both the input matrix and vector. However, in INV operations, the input matrix cannot be similarly decomposed, creating a fundamental bottleneck for achieving high precision. A recent approach addresses this by combining a low-precision, multi-level AMC inversion circuit with a high-precision, bit-sliced AMC MVM circuit, iteratively cycling between the two to solve linear systems with high accuracy [74] (Fig. 6a). Applied to MIMO detection for the first time, this method achieves high-order modulation support (e.g., 256-QAM in an 8×128 MIMO system) within just three iterations, while maintaining BER versus SNR performance comparable to a digital FP32 implementation.

Although algorithmic innovations enable high-precision solutions in massive MIMO precoding and detection, the intrinsic precision of AMC circuits (e.g., INV and GINV) still governs the convergence speed of these solutions. Enhancing AMC circuit precision remains crucial. Furthermore, AMC-based solvers exhibit inherent limitations when handling matrices with large condition numbers or unfavorable forms, which can impede MIMO performance under challenging channel conditions—an open problem that requires urgent attention.

### C. Scalability of AMC

As the number of users $K$ increases, the size of INV problems grows correspondingly. Using RRAM arrays for large-scale matrix computations faces significant challenges, including increased wiring complexity and reduced circuit stability as array size expands. Innovative strategies are therefore required. While matrix splitting is straightforward for MVM computations, it is more challenging for matrix inversion due to inter-element interactions. To address this, the BlockAMC algorithm was proposed, which decomposes the inversion of a large matrix into 2×2 block matrices, namely $A = \begin{bmatrix} M_1 & M_2 \\ M_3 & M_4 \end{bmatrix}$ [75]. For solving $Ax = b$, with $x = \begin{bmatrix} y_b \\ z_b \end{bmatrix}$ and $b = \begin{bmatrix} f \\ g \end{bmatrix}$, three INV operations of small block matrices are used alongside two MVM operations (Fig. 6b), where $M_{4s} = M_4 - M_3 M_1^{-1} M_2$. Simulations show that this approach improves precision compared with directly performing INV on a large array. Moreover, combining this block method with high-precision inversion techniques enables both scalable and precise matrix equation solving [74].

However, the block approach has limitations. Computing $M_{4s}$ is still computationally intensive and challenging to parallelize, which can affect efficiency and practical deployment. Inversion of other irregular large matrices also poses difficulties. In many MIMO applications, large sparse Gram matrices are common, leading to wasted storage in AMC-based sparse matrix computations—particularly for operations other than MVM, which cannot benefit from compression formats and SpMV optimizations. Developing compression methods for sparse INV operations thus represents a key direction for future research.

### D. Data conversion bottleneck

Although AMC shows great promise for MIMO BS signal processing, most BS algorithms are still executed in digital units, and data transfer between modules relies heavily on DACs and ADCs—either to move data from digital processors to analog units or to feed back analog results to digital units. For example, after OFDM signal modulation, the signal undergoes operations such as CP insertion and windowing before being transmitted to the RF transceiver via a DAC (Fig. 1a). These digital-to-analog (and analog-to-digital) conversions introduce significant hardware overhead in latency, power consumption, and area. To alleviate this, an analog voltage-follower scheme [50] connects AMC-based inversion units and MVM units, enabling direct analog data streaming. AMC circuits can also process OFDM baseband signals directly from or to the RF front-end, bypassing RF-side AD/DA conversions [36, 40]. While these approaches address the data conversion bottleneck, cascading analog units introduces the risk of noise amplification, which can degrade signal integrity. Potential mitigation strategies include selecting low-noise components, optimizing circuit and system layouts, and employing sparse cascading combined with digital-assisted error correction.

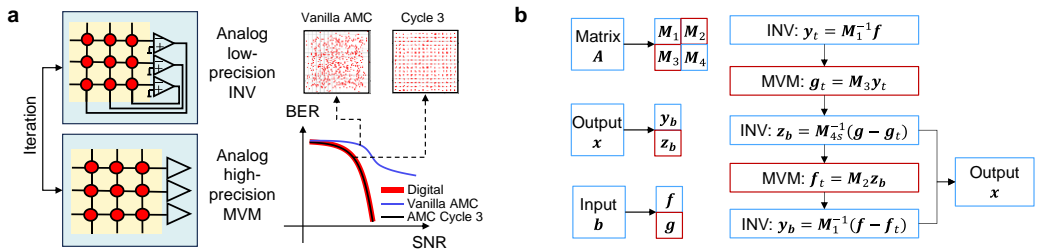

**Fig. 6.** (a) Architecture of a fully analog matrix-equation solver for MIMO detection, comprising a low-precision multi-level AMC INV circuit, a high-precision bit-sliced AMC MVM circuit, and their corresponding iterative workflow. When using the baseline AMC INV circuit, its limited computational precision leads to increased SNR loss. For high-order modulations (e.g., 256-QAM), the BER curve eventually plateaus—indicating saturated performance degradation where further SNR improvement no longer reduces the error rate. In contrast, with high-precision AMC, detection accuracy is greatly improved, and the resulting BER–SNR performance closely matches that of FP32 digital computing. (b) Architecture of BlockAMC. The large INV problem $Ax = b$ is decomposed into 3 small INV operations and 2 small MVM operations.

### E. Expanding application scenarios

Linear precoding and detection—traditional signal processing techniques—offer significant hardware deployment advantages compared to ML detection, which suffers from exponential computational complexity. However, simple linear algorithms (e.g., ZF/MMSE) still struggle with inter-user interference. To mitigate such interference, leakage-based linear precoders can be adopted [76], and the eigenvalue decomposition involved in this method can, in principle, be accelerated using AMC eigenvector circuits [13]. Previous eigenvector circuits, however, were limited to computing only the dominant eigenvector and cannot obtain the remaining eigenvectors. To overcome this limitation, an efficient AMC method has been proposed for estimating all eigenvalues [77, 78]. This approach employs a time-domain scanning technique to sweep the circuit parameter, enabling eigenvalue detection. As the scan approaches a particular eigenvalue, the nonlinear characteristics of the ridge regression circuit produce distinct voltage peaks or valleys, which reveal the eigenvalue while simultaneously outputting its corresponding eigenvector. Furthermore, with increasingly diverse antenna configurations—especially when the number of users approaches the number of antennas—AMC implementations of algorithms such as belief propagation (BP) or expectation propagation (EP) represent a promising avenue for addressing the performance limitations of linear algorithms.

Beyond these signal processing algorithms, other important applications, such as channel estimation, have not yet been fully implemented with AMC. Channel estimation can be performed via compressed sensing reconstruction, and AMC has been used to accelerate the matrix-matrix multiplication modules in algorithms like the compressed sensing local competitive algorithm (LCA) [79], achieving 1–2 orders of magnitude speedup. This advancement supports AMC deployment for channel estimation tasks.

Moreover, wireless communication signal processing extends beyond QAM/OFDM symbol manipulation to bit-stream operations, such as channel encoding (Fig. 1a). Channel coding enhances robustness against interference by introducing redundant information for error correction. Key techniques, including forward error correction (FEC) and low-density parity-check (LDPC) codes, can be implemented using AMC architectures. FEC encoding/decoding leverages MVM operations between signals and generator/decoder matrices [35]. LDPC implementations may employ ordinary differential equation (ODE) solvers constructed from linear or nonlinear AMC units with integrator-based feedback loops [80], analogous to approaches used for MIMO detection with Ising machines [81]. Realizing such complex nonlinear computations in analog circuits remains a critical research frontier.

## V Conclusion

As an emerging and promising solution, RRAM-based AMC addresses the matrix-operation–intensive demands of massive MIMO, a core technology for 5G-Advanced and 6G. Unlike digital processors constrained by computational complexity, limited throughput, and diminishing scaling benefits, AMC achieves high speed and exceptional energy efficiency by directly exploiting fundamental matrix operations—the basis of precoding, detection, DFT/IDFT, and other essential MIMO processing tasks. Despite its potential, several critical challenges still hinder large-scale deployment. The limited endurance of RRAM conflicts with frequent channel updates in time-varying environments; AMC circuits face intrinsic precision bottlenecks; large arrays encounter wiring complexity and stability issues; and much of today's research remains focused on isolated hardware components rather than system-level integration, resulting in persistent AD/DA bottlenecks and restricted application scenarios. Future advancements will rely on device–circuit–algorithm co-design, including heterogeneous architectures that couple physically analog acceleration with digital programmability. By addressing these challenges, RRAM-based AMC can fully realize the promise of massive MIMO, enabling high-performance, energy-efficient communication systems for the 6G era.

**Acknowledgements:** This work has received funding from the National Natural Science Foundation of China (62572011), the Beijing Natural Science Foundation (4252016), and the 111 Project (B18001).

**Data availability:**
Source data that support the findings of this study are available from the corresponding authors upon reasonable request.

**Code availability:**
The code used in this paper is available from the corresponding authors upon reasonable request.

**Author Contributions:** Z.S. conceived the idea for this review on RRAM-based AMC for massive MIMO signal processing. P.Z. surveyed the relevant works. P.Z. and Z.S. wrote the manuscript. Z.S. supervised the project.

**Competing interests:** The authors declare that they have no competing interests.

**Additional information**
**Correspondence** should be addressed to Z.S.